\documentclass[fleqn,usenatbib]{mnras}

\usepackage{newtxtext,newtxmath}
\usepackage[T1]{fontenc}

\DeclareRobustCommand{\VAN}[3]{#2}
\let\VANthebibliography\thebibliography
\def\thebibliography{\DeclareRobustCommand{\VAN}[3]{##3}\VANthebibliography}

\usepackage{graphicx}	
\usepackage{amsmath}	
\usepackage{lipsum}
\usepackage{subcaption}
\usepackage{xspace}
\usepackage{xfrac}
\usepackage{soul}

\newcommand{\brsfr}{\langle b\rho_{\rm SFR}\rangle}
\newcommand{\rsfr}{\rho_{\rm SFR}}
\newcommand{\nv}{\hat{\bf n}}

\def\flamingo{\texttt{FLAMINGO}\xspace}

\newcommand{\planck}{{\sl Planck}\xspace}

\newcommand{\flamfid}{{\tt L1\_m9}\xspace}
\newcommand{\flamfgas}[1]{{\tt fgas}$#1\sigma$\xspace}
\newcommand{\flamjet}{{\tt Jet}\xspace}
\newcommand{\flamjethi}{{\tt Jet\,fgas}$-4\sigma$\xspace}
\newcommand{\flammstar}{{\tt M*}$-\sigma$\xspace}
\newcommand{\flammstarhi}{{\tt M*}$-\sigma+${\tt fgas}$-4\sigma$\xspace}
\newcommand{\flamls}{{\tt LS8}\xspace}
\newcommand{\flamlshi}{{\tt LS8\,fgas}$-8\sigma$\xspace}

\title[Cosmic Infrared Background cross-correlations]{The impact of feedback and cosmology on Cosmic Infrared Background cross-correlations}

\author[S. Maleubre et al.]{
Sara Maleubre,$^{1}$\thanks{E-mail: sara.maleubremolinero@physics.ox.ac.uk}
David Alonso,$^{1}$
Reggie Dunsdon,$^{1}$
Adrien La Posta,$^{1}$
Amy Wayland,$^{1}$ and
Joop Schaye$^{2}$\\
$^1$Department of Physics, University of Oxford, Denys Wilkinson Building, Keble Road, Oxford OX1 3RH, United Kingdom\\
$^{2}$Leiden Observatory, Leiden University, PO Box 9513, NL-2300 RA Leiden, the Netherlands
}

\date{Accepted XXX. Received YYY; in original form ZZZ}

\pubyear{2025}

\begin{document}
\label{firstpage}
\pagerange{\pageref{firstpage}--\pageref{lastpage}}
\maketitle

\begin{abstract}
  We study the evolution and clustering properties of the star formation rate  density (SFR) $\rho_{\rm SFR}$ in the \flamingo suite of hydrodynamical simulations, with the aim of exploring the possibility of constraining AGN and stellar feedback prescription, as well as the effect of cosmology, using cross-correlations between the Cosmic Infrared Background (CIB) and other tracers of the large-scale structure. Specifically, we consider tomographic measurements of the bias-weighted SFR-density $\langle b\rho_{\rm SFR}\rangle$, as well as cross-correlations between the CIB and cosmic shear as a probe of the $\rho_{\rm SFR}$-matter correlation. Most of the \flamingo models calibrated against the $z=0$ galaxy stellar mass function (SMF) yield similar predictions for the distribution and clustering of star-forming systems, and thus it is difficult to derive clean conclusions regarding the impact of AGN feedback or cosmology on CIB statistics. However, we show that current measurements of $\langle b\rho_{\rm SFR}\rangle$, as well as CIB-shear cross-correlations clearly disfavour models with strong stellar feedback, as well as simulations with strong kinetic (jet-driven) AGN feedback, even under the fiducial SMF calibration. This study demonstrates the potential of CIB cross-correlations to study aspects of the physics of star and galaxy formation in a cosmological setting.
\end{abstract}





\section{Introduction}\label{sec:intro}
  Understanding how galaxies formed their stars across cosmic time is a central problem in galaxy formation and evolution \citep{1980FCPh....5..287T}. A key quantity for characterising this evolution is the star formation rate density $\rho_{\mathrm{SFR}}(z)$, which measures the total rate at which gas is converted into stars per unit of comoving volume \citep{2006A&A...451..417D}. Decades of multiwavelength observations have established a consistent picture of its evolution, whereby the SFR-density rises from the reionisation epoch towards a peak at approximately $z\sim1.9$, before declining at later times \citep{2014ARA&A..52..415M}. 

  Reconstructing this star-formation history, however, requires accounting for the substantial fraction of star formation that is obscured by dust, and was traditionally inaccessible to optical surveys such as \emph{Hubble} \citep{2001NewAR..45..601C, 2011A&A...534A..81R, 2017ApJ...838..137P}. Ultraviolet and optical radiation emitted by young stars is absorbed by interstellar dust and re-emitted at infrared wavelengths. The integrated infrared emission from unresolved dusty star-forming galaxies across cosmic history thus dominates the so-called Cosmic Infrared Background (CIB). As a result, the CIB contains an integrated record of dust-obscured star formation over a broad range of redshifts and provides a complementary probe of the cosmic star-formation history to measurements based on individually detected galaxies. In particular, analyses of CIB anisotropies have shown that the CIB can be used to constrain both the evolution of the SFR-density and the dark-matter haloes hosting dusty star-forming galaxies \citep{2018A&A...614A..39M}, which will become important in our analysis.

  The integrated two-dimensional nature of the CIB presents an important obstacle to its use in the reconstruction of the star formation history. Observations of the CIB alone cannot disentangle the contributions from star-forming sources at different redshifts without introducing additional modelling assumptions \citep{2012MNRAS.421.2832S,2021A&A...645A..40M}. This degeneracy can be broken through clustering tomography \citep{1810.00885}, by cross-correlating the CIB with tracers of the large-scale structure selected over narrow redshift bands \citep{2204.01649, 2209.05472}. Since IR-emitting galaxies trace the same underlying structure, these cross-correlations isolate the contribution from star formation associated with a corresponding epoch, allowing the redshift evolution of the CIB to be reconstructed.

  Beyond reconstructing the global star-formation history, this technique also opens the door to understanding the connection between star formation and dark-matter haloes, essential for identifying the physical processes that regulate galaxy formation. Star formation is known to be most efficient within a relatively narrow range of halo masses, while its suppression towards both lower and higher masses reflects the competing effects of stellar and AGN feedback, gas accretion and cooling, and the growth of the underlying halo population \citep{1987IAUS..117..362D, 1998A&A...331L...1S, 2006MNRAS.370..645B, 2009ApJ...696..620C}. The distribution and clustering of star-forming galaxies therefore contain information that is not captured by the integrated SFR-density alone, they encode both the halo environments in which stars are formed and the impact of baryonic processes on those environments. 

  As mentioned above, the anisotropies of the CIB provide a particularly useful avenue for accessing this information. Since dusty star-forming galaxies occupy the same large-scale structure traced by matter, cross-correlations of the CIB with other cosmological tracers can isolate the star-forming signal over different redshifts and physical scales. On large scales, these measurements probe the connection between the SFR-density and the clustering of the haloes hosting it, allowing the evolution of the bias-weighted SFR-density $\brsfr$ to be reconstructed tomographically \cite{2206.15394}. On smaller scales, cross-correlations with gravitational lensing provide access to the spatial relation between star formation and the total matter distribution within and around haloes \cite{2209.05472}. CIB cross-correlations can therefore probe complementary aspects of the SFR--halo-connection, from its cosmic evolution to is non-linear clustering.

  Interpreting these measurements requires understanding how uncertain baryonic processes modify the distribution of star formation. Feedback from stars, supernovae, and active galactic nuclei (AGN) regulates the availability of star-forming gas, and can redistribute baryons over halo scales, potentially modifying both the amount of star formation associated with haloes of a given mass and its spatial relation to the underlying matter distribution \citep{2006MNRAS.365...11C,2012MNRAS.421.3522H,2012MNRAS.427.2998,2013ApJ...776....1K}. Hydrodynamical simulations provide a natural laboratory in which to investigate these effects. In particular, suites containing controlled variations of their baryonic prescriptions allow us to determine which properties of the star-forming population are robust and which retain sensitivity to the uncertain implementation and calibration of feedback \citep[e.g.][]{2010MNRAS.402.1536S,2012ASSL..378...83C,2015MNRAS.450.1937C, 2510.20897,2026MNRAS.548ag300C}.

  In this work, we use the \flamingo cosmological hydrodynamical simulations \citep{SchayeEtal2023, KugelEtal2023,2026A&C....5701159H} to investigate the impact of baryonic feedback on the connection between the CIB, star formation and the large-scale structure (LSS). The \flamingo simulation set is particularly suited to this purpose, since it features simulations sharing the same cosmology and initial conditions while varying the calibration and implementation of stellar and AGN feedback, as well as variations in the underlying cosmology. This allows us to compare the predictions from these different models against existing measurements of CIB cross-correlations with various LSS tracers in a quantitative manner.

  This paper is structured as follows. The theoretical framework for CIB--cross-correlations can be found in Section~\ref{sec:theory}, while a summary description of the simulations can be found in Section~\ref{sec:sims}. Then, in Section \ref{sec:res}, we use the different \flamingo models to study how feedback modifies the contribution to the total SFR-density from different halo masses, as well as the resulting SFR-matter power spectrum. We then compare these statistics against two complementary observables: measurements of the evolution in the bias-weighted SFR-density from CIB tomography, and cross-correlations between the CIB and weak gravitational lensing data. We conclude in Section~\ref{sec:conc}.

\section{CIB cross-correlations}\label{sec:theory}
  \subsection{CIB and star formation}\label{ssec:theory.sfr}
    The CIB is dominated by the infrared emission from dust particles, heated by the UV radiation emitted by massive, short-lived stars \citep{1967ApJ...147..868P,2001ApJ...550....7K}. The infrared emissivity thus traces the local rate at which new stars are formed. This is typically quantified by the empirical relationship between the total infrared luminosity $L_{\rm IR}$ and the SFR of a given source:
    \begin{equation}
      {\rm SFR}=K\,L_{\rm IR},
    \end{equation}
    where $K=1.0\times10^{-10}M_\odot\,{\rm yr}^{-1}\,L_\odot^{-1}$ is the calibration constant corresponding to a Chabrier initial mass function \citep{2003PASP..115..763C}, with $L_{\rm IR}$ integrated over the wavelength range 8-1000\,$\mu{\rm m}$ \citep{astro-ph/9807187,2012ARA&A..50..531K}. Note that this approximate relation implicitly assumes that the obscured SFR fraction is independent of the SFR itself, which is likely incorrect in detail. Additionally, AGN are also a source of dust heating, and a potentially important contribution to the clustering properties of the CIB. Although we ignore this contribution here, it constitutes a potentially interesting avenue to study AGN feedback through CIB cross-correlations.

    The specific intensity of the CIB at a frequency $\nu$ can then be calculated by adding the infrared emission of all star-forming regions along a given line of sight:
    \begin{equation}
      I_\nu(\nv)=\int d\chi\,\chi^2\,\frac{S^{\rm eff}_\nu(z)}{K}\,\rho_{\rm SFR}(\chi\nv).
    \end{equation}
    Here $\rho_{\rm SFR}({\bf x},z)$ is the comoving star formation rate density at comoving coordinates ${\bf x}$ and redshift $z$, $\chi$ is the radial comoving distance, and $S^{\rm eff}_\nu(z)$ is the mean flux of infrared sources at redshift $z$ normalised to one solar luminosity $L_\odot$.

    As we discuss below, cross-correlations between maps of the CIB and tracers of the large-scale structure lying at a particular range of redshifts $\sim\bar{z}$ are only sensitive to the contributions to $I_\nu(\nv)$ from sources at those redshifts. The CIB cross-correlation can thus be separated from that of other components emitting at the same frequencies (e.g. the thermal Sunyaev-Zel'dovich effect) by exploiting its distinct spectrum at that redshift $S_\nu^{\rm eff}(\bar{z})$ \citep[see e.g.][for further details]{2026arXiv260304269L,2609.31774}. Thus, CIB cross-correlations are effectively sensitive to the projected star formation rate density,
    \begin{equation}\label{eq:sigma_sfr}
      \Sigma_{\rm SFR}(\nv)\equiv\int d\chi\,\chi^2\,\rho_{\rm SFR}(\chi\nv,z),
    \end{equation}
    with specific cross-correlations picking up only the contribution to $\Sigma_{\rm SFR}$ from the redshift range covered by the other LSS tracer. This can be seen easily by examining the structure of the angular cross-power spectrum between two projected LSS tracers in the Limber approximation \citep{1954ApJ...119..655L,0809.5112}:
    \begin{equation}\label{eq:cl_limber}
      C_\ell^{uv}=\int\frac{d\chi}{\chi^2}\,q_u(\chi)\,q_v(\chi)\,P_{UV}\left(k=\frac{\ell+1/2}{\chi},z\right).
    \end{equation}
    Here $q_u$ is the radial kernel of tracer $u(\nv)$, with $U({\bf x},z)$ being the three-dimensional tracer of the LSS it is connected to:
    \begin{equation}
      u(\nv)=\int d\chi\,q_u(\chi)\,U(\chi\nv,z),
    \end{equation}
    and $P_{UV}(k,z)$ is the three-dimensional power spectrum between $U$ and $V$. In this formalism, the radial kernel for $\Sigma_{\rm SFR}$ is simply $q_{\rm SFR}(\chi)=4\pi\chi^2$, with its three-dimensional counterpart $\rho_{\rm SFR}$. As can be seen in Eq. \ref{eq:cl_limber}, the cross-correlation is only sensitive to the properties of the tracers involved in the range of redshifts in which both radial kernels overlap.

    In this work we will be interested in cross-correlations with two types of tracers: the overdensity of galaxies in a given bin of redshifts, and observed shapes of galaxies perturbed by weak gravitational lensing (cosmic shear). These will allow us to explore two aspects of $\rho_{\rm SFR}$: the evolution in the mean cosmic SFR-density, and its small-scale clustering properties (or, equivalently, the clustering of star forming galaxies). We describe each of these probes in the next two sections.

  \subsection{Galaxy cross-correlations and tomography}\label{ssec:theory.tomo}
    Galaxies are a local tracer of the large-scale structure (i.e. they are sensitive to the clustering of matter at the redshifts we observe them). Specifically, the angular overdensity of galaxies $\delta_g$ is related to their three-dimensional overdensity $\Delta_g$ in terms of their redshift distribution $p(z)$:
    \begin{equation}
      \delta_g(\nv)=\int d\chi\,H(z)\,p(z)\,\Delta_g(\chi\nv,z),
    \end{equation}
    where $H(z)$ is the expansion rate\footnote{Note that we use natural units with $c=1$ throughout.}. One can typically select galaxies in a relatively compact range of redshifts, in which case the radial kernel $q_g(\chi)\equiv H(z)\,p(z)$ is narrowly peaked around the mean distance/redshift to the sources.
    
    Cross-correlations between galaxies and the CIB are therefore sensitive to the galaxy-SFR power spectrum, $P_{g{\rm SFR}}(k,z)$ at the redshift of the galaxies. On small scales ($k\gtrsim 0.3\,{\rm Mpc}^{-1}$) $P_{g{\rm SFR}}(k,z)$ is sensitive to the detailed clustering of the target galaxies, including the non-trivial small-scale correlations with the star-forming galaxies that dominate the CIB emission. This is in general very difficult to model at sufficient accuracy. Furthermore, reliably extracting this measurement on small scales from multi-frequency data is especially complicated, since the cross-correlation is additionally sensitive to the specific (and typically unknown) infrared spectral properties of the target galaxy sample. These complications disappear if we instead target only large scales: in this regime both galaxies and the CIB/SFR-density trace the same large-scale structure. The scale dependence of the cross-correlation is characterised by the known matter power spectrum, and its amplitude depends only on the global properties of infrared/star-forming galaxies, and the large-scale bias of the target galaxies. For the same reason, on these large scales the CIB is governed by the global average spectrum of all infrared sources, simplifying the problem of component separation \citep{2026arXiv260304269L}. More concretely, on large scales
    \begin{equation}
      P_{g{\rm SFR}}(k,z)\simeq b_g\brsfr\,P_{mm}(k,z),
    \end{equation}
    where $P_{mm}(k,z)$ is the matter power spectrum, $b_g$ is the linear bias of the target sample, and $\brsfr$ is the bias-weighted average SFR-density:
    \begin{equation}
      \brsfr\equiv \int dM\,n(M)\,b_h(M)\,{\rm SFR}(M),
    \end{equation}
    where $n(M)$ and $b_h(M)$ are the halo mass function and the linear halo bias, and ${\rm SFR}(M)$ is the mean star formation rate in haloes of mass $M$. Combining this with a measurement of the large-scale galaxy auto-spectrum $P_{gg}(k)\simeq b_g^2 P_{mm}(k)$ allows us to measure $b_g$ and $\brsfr$ separately. Targetting galaxy samples at different redshift thus allows us to reconstruct the cosmic star formation history through $\brsfr$.

    A method was developed and tested in \cite{2026MNRAS.545f2125M} to estimate $\brsfr$ (and equivalent quantities for other LSS tracers) from galaxy cross-correlations, such that the resulting measurements are independent of the clustering properties of the particular galaxy sample used. The estimator includes additional terms to the naive linear bias model above that absorb the impact of scale-dependent and stochastic bias on mildly non-linear scales. 

    This method was incorporated in the multi-frequency power spectrum-based component separation framework developed in \cite{2026arXiv260304269L} to extract and analyse the cross-correlations of multiple extragalactic components \citep[see e.g.][for similar approaches]{2006.14650,2504.05384}. In Section~\ref{ssec:res.tomography} we will use the method of \cite{2026MNRAS.545f2125M} to extract $\brsfr$ as a function of redshift from the different \flamingo simulations, and compare the result with the measurements of \cite{2026arXiv260304269L} using \planck data and a variety of galaxy samples covering redshifts $z\lesssim 2.7$ .

  \subsection{Cross-Correlations with weak-lensing}\label{ssec:theory.wl}
    Cosmic shear probes the fluctuations in the overall density of matter through the distortion the associated gravitational potential field causes in the observed shapes of galaxies \citep{astro-ph/9912508}. At leading order, cosmic shear is a pure $E$-mode spin-2 field that traces the matter overdensity field $\delta_m$ from the source redshift to $z=0$:
    \begin{equation}
      \gamma^E(\nv)=\int d\chi\,q_\gamma(\chi)\,\delta_m(\chi\nv,z),
    \end{equation}
    where the \emph{lensing kernel} $q_\gamma$ is
    \begin{equation}\label{eq:q_gamma}
      q_\gamma(\chi)=\frac{3H_0^2\Omega_m}{2a}\chi\int_{z(\chi)}^\infty dz'\,p(z)\,\frac{\chi(z')-\chi}{\chi(z')}.
    \end{equation}
    Here $H_0\equiv H(0)$ is the Hubble constant, $\Omega_m$ is the fractional abundance of matter, $a\equiv1/(1+z)$ is the scale factor, and $p(z)$ is the redshift distribution of sources.
    
    The cross-correlation between the CIB and cosmic shear is thus sensitive to the matter-SFR power spectrum
    \begin{equation}
      P_{m\,{\rm SFR}}(k)\propto \langle \rho_{\rm SFR}({\bf k})\delta_m^*({\bf k})\rangle
    \end{equation}
    As in the case of $P_{g{\rm SFR}}(k)$, on large-scales $P_{m{\rm SFR}}(k)$ probes the bias-weighted average SFR denstiy. On the other hand, on small scales this probe is sensitive to the abundance and distribution of star-forming galaxies within dark matter haloes, and their non-linear clustering properties on intermediate scales. Importantly, and unlike in the case of $P_{g{\rm SFR}}(k)$, this observable depends on the relation between the global population of star-forming galaxies and the overall matter inhomogeneities, and not on the clustering properties of a specific galaxy sample, or their complex relation with star-forming sources residing in the same haloes. This makes modelling $P_{m{\rm SFR}}(k)$ on small scales qualitatively simpler, removing the need to fully describe the properties of specific galaxy samples. 
    
    This is important in the case of cosmic shear cross-correlations: the lensing kernel has support from $z=0$ to the source, peaking at approximately half that distance. Thus, a given angular scale $\ell$ receives non-negligible contributions from a wide range of comoving scales $k$. This means that it is often difficult to separate large, linear scales from small, non-linear ones (the latter introduced by the low-redshift tail of $q_\gamma(\chi)$) where the modelling of small-scale clustering is crucial.


\section{Simulations}\label{sec:sims}
  We use the \flamingo suite of cosmological hydrodynamical simulations \citep{SchayeEtal2023, KugelEtal2023}, designed to model large-scale structure while simultaneously resolving the galaxy populations responsible for baryonic effects. The simulations were performed with the \textsc{Swift} code \citep{SchallerEtal2024b}, using the SPHENIX implementation of smoothed-particle hydrodynamics \citep{BorrowEtal2022}. Unresolved physical processes are treated with subgrid prescriptions, including radiative cooling and heating \citep{2020MNRAS.497.4857P}, star formation and evolution \citep{2008MNRAS.383.1210S, 2009MNRAS.399..574W}, black hole growth \citep{2009MNRAS.398...53B, 2022MNRAS.516..167B}, feedback from young stars and supernovae \citep{2022MNRAS.514..249C,2023MNRAS.523.3709C}, and AGN feedback \citep{2009MNRAS.398...53B, 2022MNRAS.516.3750H} 

  In this work, we focus on the intermediate-resolution set of simulations. These have a box size of $(1\,\mathrm{Gpc})^{3}$ (${\tt L1}$) and baryonic particle masses of $m_{\mathrm{gas}}=1\times10^9M_{\odot}$ (${\tt m9}$). All baryonic variations considered here share the same initial conditions and baseline D3A cosmology, with $\Omega_{\rm m}=0.306$, $\Omega_{\rm b}=0.0486$, $h=0.681$, $n_s=0.967$, $A_s=2.099\times10^{-9}$ and a total neutrino mass of $0.06\,{\rm eV}$. The use of identical initial conditions and cosmological parameters allows differences between these simulations to be associated directly with changes in their baryonic modelling. In addition, we also study the two \emph{low}-$\sigma_8$ ($\sigma_8=0.760$) simulations available, one with the \emph{fiducial} feedback (\flamls) and another one with the \emph{strongest thermal-AGN} feedback (\flamlshi).

  A defining feature of \flamingo is that the parameters controlling stellar and AGN feedback are calibrated against observational data. The fiducial \flamfid model is calibrated to reproduce the low-redshift galaxy stellar mass function (SMF) in the range of masses $\rm 10^{9.92} M_{\odot}<{\it M}_{\star}<10^{11.5} M_{\odot}$, and the observed gas fractions $f_{\rm gas}$ of galaxy groups and clusters with $\rm 10^{13.5} M_{\odot}<{\it M}_{500c}<10^{14.36}M_{\odot}$. The suite of modified feedback models shifts the calibration data in relation to their observational errors, allowing also for variations in the physical implementation of AGN feedback. Specifically, two models were used: in the fiducial ``thermal injection'' model, heat is deposited by the central black hole, raising the temperature of nearby gas particle by a given amount $\Delta T_{\rm AGN}$. In the alternative ``kinetic injection'' or ``${\tt Jet}$'' model, AGN-driven jets provide a ``kick'' to pairs of gas particles within the central black hole's kernel with a velocity $v_{\rm jet}$ along the spin axis. $\Delta T_{\rm AGN}$ and $v_{\rm jet}$ are treated as free parameters. In turn, stellar feedback is parametrised in terms of the fraction of the energy injected by stars and supernovae that couples with the inter-stellar medium, $f_{\rm SN}$, as well as the velocity of the galactic wind $\Delta v_{\rm SN}$. Crucially, model variations in \flamingo are not expressed only as changes in these subgrid parameters: each simulation is defined by the relative shift in the target observables (SMF and $f_{\rm gas}$), with each new calibration triggering a multiple-parameter adjustment. These variations provide a controlled way of exploring the sensitivity of cosmological observables to uncertainties in baryonic physics observables.

  The simulations considered in this work can be broadly divided into three classes:
  \begin{itemize}
    \item Variations in $f_{\rm gas}$: the observational gas-fraction relation is shifted by a given number of standard deviations, $\pm X\sigma$, before recalibrating the subgrid parameters. These simulations are labelled as \flamfgas{\pm X}. Lower target gas fractions require feedback capable of removing a larger fraction of gas from groups and clusters and therefore broadly correspond to increasingly efficient AGN-driven gas ejection. These variations are intentionally broader than the observationally preferred range and are used to span a conservative range of plausible baryonic effects.
    \item Modification to the stellar-mass calibration, $M_\star$: the observational stellar masses entering the SMF are shifted downwards by $0.14$ dex (labelled as \flammstar), with the rest of parameters recalibrated to reproduce this modified SMF. This typically translates into a stronger stellar-feedback, making them particularly useful for testing how changes in the efficiency with which galaxies build their stellar mass propagate into the cosmic star-formation history.
    \item Changes in the physical implementation of AGN-feedback through jets (labelled as \flamjet). These models allow us to investigate whether observables sensitive to star formation depend only on the calibrated baryonic content of haloes or additionally retain sensitivity to the mechanism through which AGN feedback couples to the gas.
  \end{itemize}
  We not only explore these models in isolation, but also their combination in
  \flammstarhi (shifts in both the SMF and $f_{\rm gas}$ observables) and \flamjethi (shifts in $f_{\rm gas}$ within the kinetic injection model). 

  Haloes and galaxies are identified in the simulation with an updated version \citep{Forouhar-MorenoEtal2025} of the HBT (Hierarchical Bound Tracing, \cite{HanEtal2018}) algorithm, while their properties are computed with a \flamingo-specific tool called SOAP \citep[Spherical Overdensity and Aperture Processor,][]{2507.22669}. These catalogues provide the halo masses and total halo SFRs used to study the halo-mass distribution of the SFR-density and its evolution across the different baryonic models.

  The \flamingo simulation data products are available through the \flamingo database\footnote{\href{https://flamingo.strw.leidenuniv.nl/data.html}{https://flamingo.strw.leidenuniv.nl/data.html}} \citep{2026A&C....5701159H}. 
  


\section{Results}\label{sec:res}
  \subsection{SFR across halo mass}\label{ssec:res.SFR_mass_rel}
    \begin{figure*}
      \raggedright
      \includegraphics[width=0.945\linewidth]{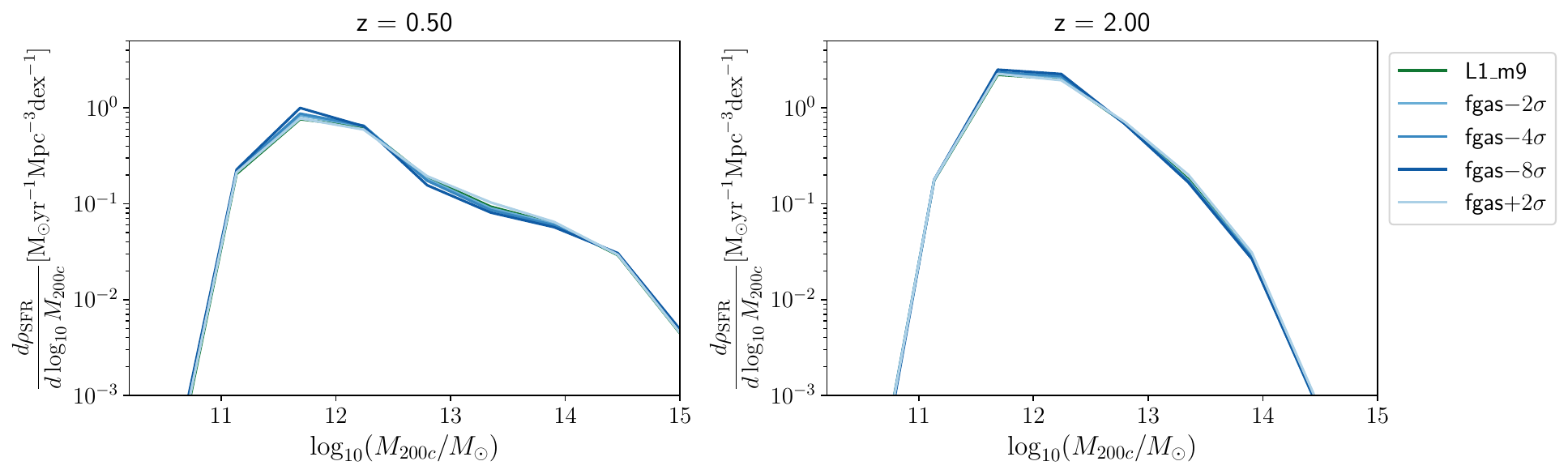}
      \includegraphics[width=\linewidth]{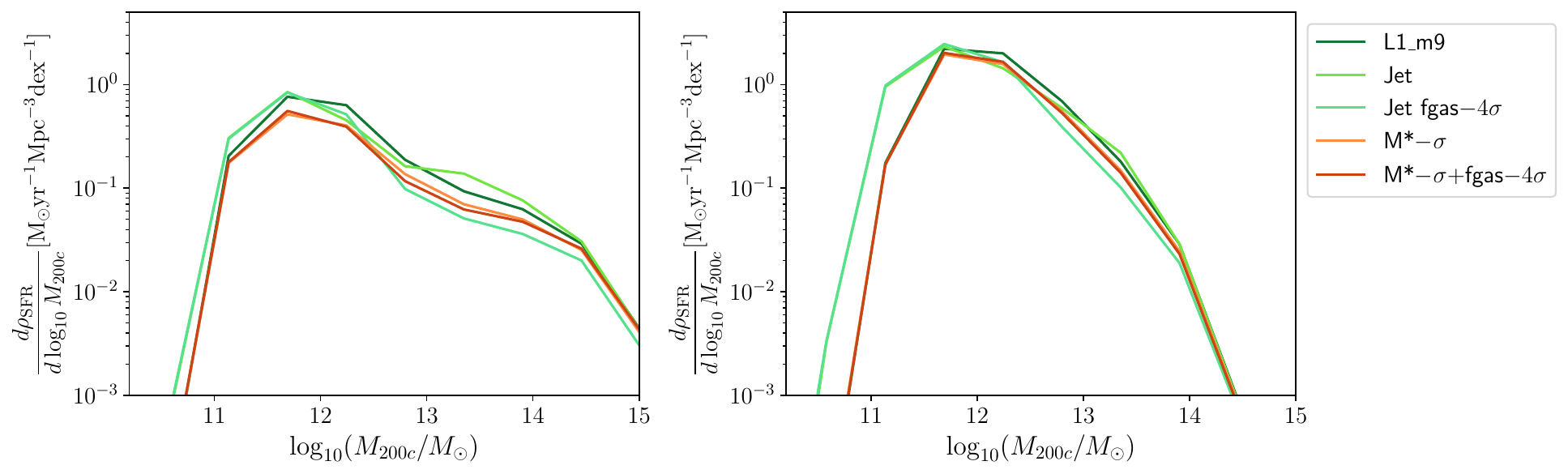}
      \caption{Contribution to the SFR-density per logarithmic halo-mass interval for the different feedback models at redshifts $z=0.5$ and $z=2.0$ (left and right panels, respectively). The top panels show the predictions from the different \flamfgas{\pm N} models, while the bottom panels show the results from the \flamjet and \flammstar variations, in both cases compared with the fiducial \flamingo simulation.}
      \label{fig:rhosfrperdex}
    \end{figure*}
    To aid in the interpretation of the observational data presented in subsequent sections, we start by quantifying how haloes of different mass contribute to the total star formation budget in the different \flamingo simulations, with the aim of understanding how different implementations of baryonic physics affect the SFR history and distribution. We first want to investigate the halo masses that dominate the star formation rate density, and how this distribution is affected by variations in the baryonic physics implemented in \flamingo. We estimate the contribution to the total SFR-density from haloes in a logarithmic mass interval of width $\Delta \log_{10}M_{200c}$, by summing the SFR associated with all haloes within the bin:
    \begin{equation}
      \frac{{\rm d}\rsfr}{{\rm d}\log_{10}M_{200c}}\simeq \frac{\sum_{h\in {\rm bin}}{\rm SFR}_h}{V_{\rm sim}\Delta\log_{10}M_{200c}}.
    \end{equation}

    Figure~\ref{fig:rhosfrperdex} shows the resulting halo-mass-binned SFR-density at two redshifts,  $z=0.5$ and $z=2.0$, spanning the range covered by current CIB cross-correlation measurements. At both redshifts, the SFR-density is strongly concentrated in a relatively narrow range of halo masses, with the maximum peaking at around $(M_{200c}\sim10^{12}\,{\rm M_\odot})$. This is an unsurprising result, as star formation is known to be inefficient at both the low- and high-mass ends \citep{2013A&A...557A..66B}. At low masses, the shallow gravitational potential makes it difficult for galaxies to retain their gas as stellar feedback from supernovae can efficiently heat and expel their already small gas reservoirs, producing a steep decline in SFR-density \citep{2003MNRAS.343..249S}. The limited mass resolution of the \flamingo simulations likely also plays a role in this regime. At the opposite mass end, gas cooling times become significantly longer, and star formation is affected by AGN feedback \citep{2005MNRAS.363....2K}. In addition, haloes become significantly rarer as we move to higher masses, which means that even if an individual system contains a meaningful amount of star-forming gas, their overall contribution is small with respect to the total SFR budget. Thus the drop at high masses reflects a combination of suppressed star formation efficiency and the rapidly declining halo abundance. This behaviour agrees with that found in the higher-resolution COLIBRE simulations \citep[see Fig. 6 in ][]{2509.07960}.

    Once this general behaviour has been established, we can look at small variations dependent on the feedback model. At both redshifts we see that the two {\tt Jet} models show a distinct enhancement of the contribution from lower-mass haloes, a tendency that increases with redshift. In addition, at higher masses we can also appreciate a different behaviour between the \flamjet and \flamjethi simulations. While the former experiences an enhanced star formation with respect to the fiducial \flamingo model, the latter is considerably suppressed at both redshifts. This is in contrast to the impact of the thermal injection model as a means to modify the predicted gas fractions, where the effect is almost negligible. This behaviour may not be directly related to the kinetic nature of the AGN feedback model in these simulations. A key difference in the \flamjet and \flamjethi model with respect to the other {\tt m9\xspace} resolution \flamingo simulations is the implementation of a more efficient black hole repositioning scheme, which results in more efficient quenching and AGN feedback \citep[see Fig. 14 in ][]{SchayeEtal2023}. Note also that the highest galaxy masses ($\log M_*/M_\odot > 11.5$) were not used in the SMF calibration of \flamingo, which gives the different feedback models more freedom in the high-mass regime.

    We can also observe that the \flammstar models suppress the SFR overall, and particularly around the peak halo mass. This is not surprising, given the requirement to reproduce lower galaxy stellar masses and thus shift the SMF at $z=0$. From the subgrid parameters of these simulations \citep[see][]{SchayeEtal2023}, this is largely achieved by strengthening the stellar feedback parameters and, to a lesser extent, the impact of thermally-driven AGN feedback. It is also noticeable that this SFR suppression becomes increasingly prominent at lower redshifts. Since the modified SMF constitutes a low-redshift calibration target, this behaviour might suggest that the effects of the recalibrated feedback model build up over the star-formation history, producing progressively larger differences in the galaxy population towards late times. 

    Finally, as anticipated before, when looking across the different \flamfgas{\pm N} models, we barely observe any variations between them. Their impact on the halo-mass-binned SFR-density is therefore comparatively small, despite corresponding to substantial changes in the amount of gas retained by massive haloes. This difference in behaviour could be explained by the differences in calibration targets. While \flammstar models act on the gas supply and star-formation of galaxy-scale systems (around the $\sim10^{12}{\rm M_{\odot}}$ peak), the \flamfgas{\pm N} models change the hot gas fraction of more massive systems, which contribute relatively little to the total SFR-density. This can also be seen in the negligible differences between \flammstar and \flammstarhi.

\begin{figure}
      \centering
      \includegraphics[width=\linewidth]{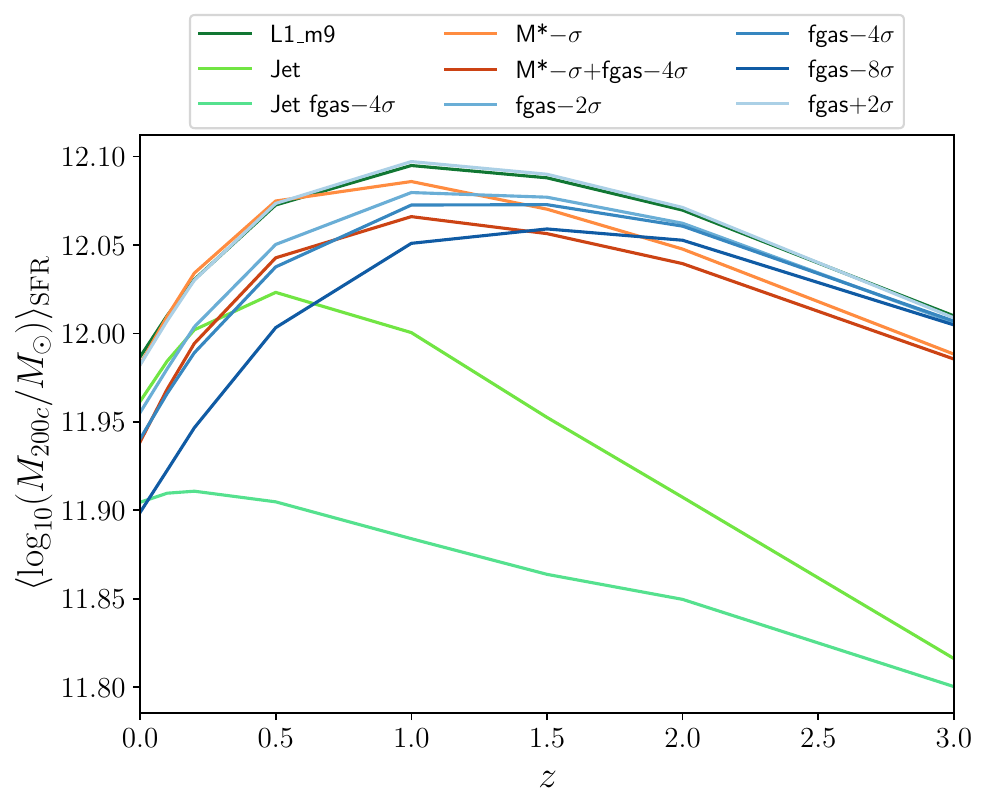}
      \caption{SFR-weighted mean logarithmic halo mass for the different feedback models in \flamingo.}
      \label{fig:Mh_SFR}
    \end{figure}

    To further quantify how these feedback variations modify the SFR distribution across halo masses, we considered the SFR-weighted mean logarithmic halo mass
    \begin{equation}
      \left\langle \log_{10} M_{200c} \right\rangle_{\rm SFR}=
      \frac{
        \sum_i {\rm SFR}_i\,\log_{10} M_{200c,i}
      }{
        \sum_i {\rm SFR}_i
      }.
    \end{equation}
    This quantity summarizes the halo-mass distribution shown in Figure~\ref{fig:rhosfrperdex}, providing a compact measure of the typical mass scale of haloes contributing to the SFR budget. Figure~\ref{fig:Mh_SFR} shows the evolution of the SFR-weighted mean halo mass as a function of redshift. For most of the \flamingo models, the characteristic halo mass remains remarkably stable, close to the reported peak at $(M_{200c}\sim10^{12} {\rm M_\odot})$, and a very mild redshift evolution is present (likely reflecting the balance between the growth of efficient star-forming structures and the increasing suppression of star formation in massive systems towards late times). 
    As predicted, although the \flammstar models show a systematically lower SFR distribution as a function of mass, their SFR-weighted mean halo masses remain relatively close to the fiducial model. The effect of the modified stellar-mass calibration therefore appears to be predominantly a suppression of star formation within approximately the same halo population, rather than a substantial redistribution towards different halo masses. Similarly, the \flamfgas{\pm N} variations produce only small changes in the SFR-weighted mean halo mass, as expected from Fig. \ref{fig:rhosfrperdex}.

    On the other hand, the two {\tt Jet} models show a much stronger deviation from the fiducial behaviour. Their SFR-weighted mean halo mass decreases faster and more significantly at high redshifts than the rest. This behaviour is consistent with the redistribution of SFR towards lower-mass haloes seen in Figure~\ref{fig:rhosfrperdex}.

    This redistribution of SFR is relevant in the context of tomographic measurements of $\brsfr$. Variations in this quantity may be caused by either an overall enhancement or suppression of the total SFR budget (e.g. as in the case of the \flammstar simulations), as well as by changes in the mass dependence of the SFR coupled with the mass-dependent halo bias (as should likely be the case in the two {\tt Jet} models).

  \subsection{Tomography}\label{ssec:res.tomography}
    We next investigate the halo-bias-weighted SFR-density, $\brsfr$, focusing on the predictions from the different \flamingo feedback models, and compare them to current measurements of this quantity. To extract $\brsfr$ from the simulations we used the methodology developed in \citet{2026MNRAS.545f2125M}, ensuring that the measurements are not affected by non-linear biasing. This methodology was applied to measurements of the cross-correlation of galaxies and multi-frequency CMB maps in \citet{2026arXiv260304269L} to extract measurements of $\brsfr$ over the redshift range $z\lesssim1$. We will compare the \flamingo predictions against these data, in addition to newer measurements using high-redshift quasars from the {\sl Quaia} sample \citep{2306.17749}. We made these measurements following the methodology described in \citet{2026arXiv260304269L}, using the two higher-redshift bins of the {\sl Quaia} catalog described in \citet{2402.05761}. These measurements expand the redshift range to $z\lesssim2.7$, reaching the peak of the star formation history, and measuring $\brsfr$ above $z\simeq1.5$ for the first time \citep[see also][]{2206.15394}. The new measurements are:
    \begin{eqnarray}
      &\brsfr_{z=1.6}= 0.217 \pm 0.015\,M_\odot\,{\rm yr}^{-1}{\rm Mpc}^{-3},\\
      &\brsfr_{z=2.7} = 0.371\pm0.071\,M_\odot\,{\rm yr}^{-1}{\rm Mpc}^{-3}.
    \end{eqnarray}

    \begin{figure*}
      \centering
      \includegraphics[width=\linewidth]{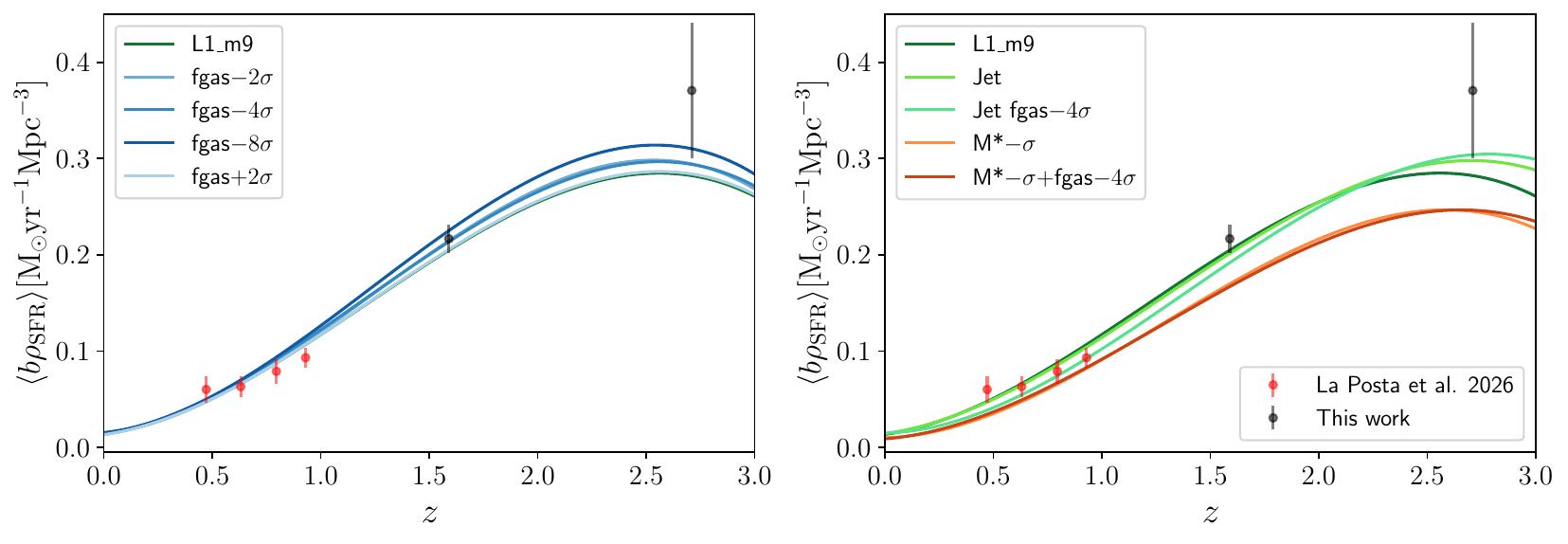}
      \caption{Halo-bias weighted SFR-density as a function of redshift for different feedback models. We have separated the purely thermal AGN feedback models (left panel) from the jet-based AGN feedback and strong stellar feedback models (right panel), with the fiducial $L1\_m9$ model present in both for easy comparison. We also plot observational results from \citet{2026arXiv260304269L} and those obtained from the cross correlation of \emph{Planck} with {\sl Quaia} quasar catalogue (black).}
      \label{fig:rhoSFR}
    \end{figure*}

    Figure~\ref{fig:rhoSFR} shows the redshift evolution of $\brsfr$, divided into the same two families of feedback models used in Fig. \ref{fig:rhosfrperdex}. In the left panel, the \flamfgas{\pm N} models display relatively small variations in $\brsfr$ with respect to the fiducial model at low redshifts, becoming more prominent at higher redshifts. This behaviour is consistent with the weak dependence described in the previous section. Interestingly, we find that the simulations with suppressed cluster gas fractions (\flamfgas{-N}) display marginally \emph{larger} $\brsfr$ at high redshifts. This can be understood as follows: to compensate for the stronger AGN feedback in these simulations, the stellar feedback subgrid parameters are weakened in order to reproduce the SMF at $z=0$, allowing for slightly larger SFRs at higher redshifts. In spite of these differences, all models are largely in agreement with the data, with deviations of at most $\sim1.5$-$2\sigma$.
    
    The right panel shows the predictions from the \flamjet and \flammstar models, where larger deviations are observed with respect to the fiducial simulation, in line with the results show in the previous section. The \flammstar models recover a lower value of $\brsfr$ consistently at all redshifts, in accordance with the general suppression of SFR as a function of mass observed in Fig. \ref{fig:rhosfrperdex}. These predictions are in clear disagreement with the data, particularly at intermediate redshifts $z\simeq1.5$. In turn, the \flamjet and \flamjethi models display a qualitatively different response. Below $z\simeq2$, the \flamjet model recovers broadly the same prediction as the fiducial model, while larger $\brsfr$ values are predicted at higher redshifts. This is due to the enhanced SFR at low halo masses of this model discussed in the previous section (see Fig. \ref{fig:rhosfrperdex}), which lead to an overall increase in the total SFR budget \citep[as seen also in E.g. Fig. 13 of ][]{SchayeEtal2023}. In turn, the \flamjethi model predicts a lower value of $\brsfr$ at $z\lesssim 2$, while reproducing the same enhancement at higher redshifts. The low-redshift behaviour of \flamjethi is in contrast with the thermal injection models, which instead predict marginally higher $\brsfr$ values. As discussed earlier, this is likely caused by more efficient black hole repositioning scheme implemented in the two {\tt Jet} models, leading to stronger quenching. In spite of these differences, the observational data is not precise enough to clearly favour or disfavour either of these models with respect to the fiducial simulation.

    As we have seen, tomographic measurements of $\brsfr$, quantifying the abundance and large-scale clustering of star-forming systems, can be used to place constraints on different stellar feedback scenarios, with the \flammstar models showing a clear disagreement with the data. Further insights may be gained by exploring the small-scale clustering properties of the star formation density, as quantified by the matter-$\rsfr$ power spectrum, measurable in CIB-shear cross-correlations.

  \subsection{The matter-SFR cross-power spectrum}\label{ssec:res.mSFR_PS}
    Having characterised the large-scale behaviour of the SFR field through $\brsfr$, we now investigate its small-scale dependence using the matter-SFR cross-power spectrum, $P_{m\mathrm{SFR}}(k)$.
    
    \begin{figure*}
      \centering
      \includegraphics[width=\linewidth]{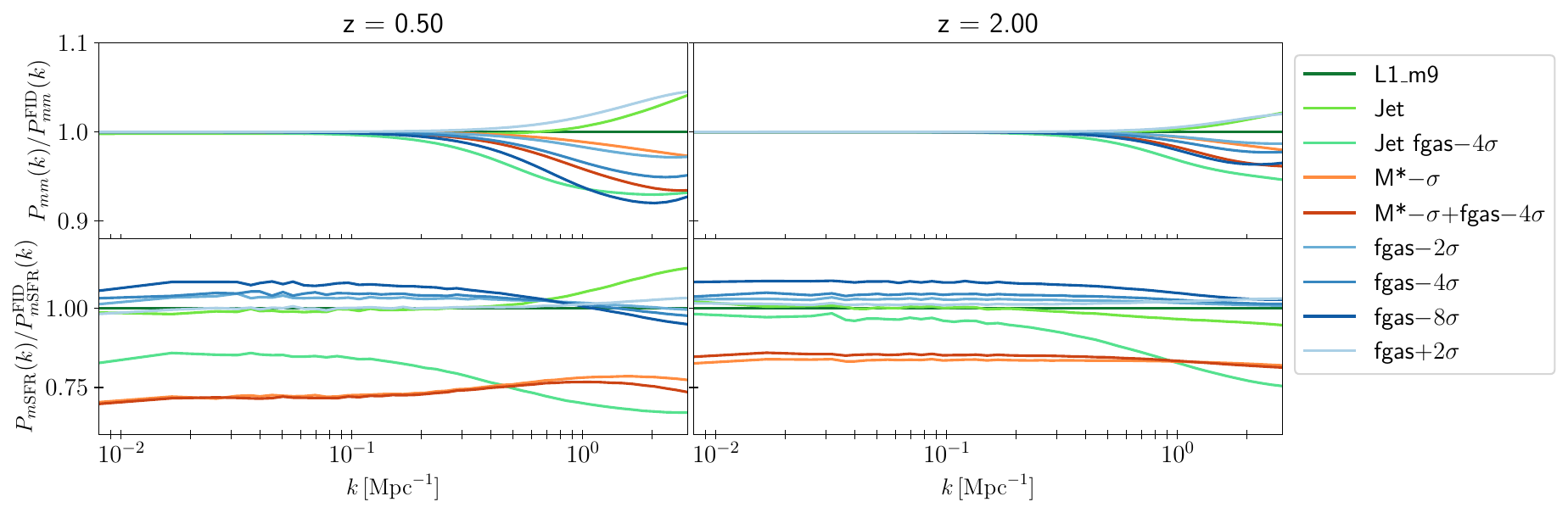}
      \caption{Ratio of power spectra from the different \flamingo simulations with respect to the fiducial model \flamfid. Results are shown for the matter auto-power spectra (upper panels) and the matter-SFR cross-power spectra (lower panels). Results are shown at redshifts $z=0.5$ and 2 (left and right, respectively).}\label{fig:PSmSFR}
    \end{figure*}

    The lower panels in Figure~\ref{fig:PSmSFR} shows the ratio of the matter-$\rsfr$ power spectrum, $P_{m\mathrm{SFR}}(k)$, in different simulations with respect to the fiducial (\flamfid) \flamingo model at redshifts $z=0.5$ and $2.0$. On sufficiently large scales the matter overdensity is insensitive to baryonic feedback, thus the differences seen here in $P_{m\mathrm{SFR}}(k)$ reflect the corresponding differences in $\brsfr$ discussed in the previous section. Towards smaller, non-linear scales, baryonic feedback modifies the distribution and clustering of star-forming systems, as well as the underlying matter distribution. It therefore provides additional information, beyond $\brsfr$, regarding the coupled response of matter and star formation to feedback, probing the small-scale dependence relevant to weak lensing analyses. The observed variations in $P_{m\mathrm{SFR}}(k)$ can be due to three distinct effects, or a combination thereof.
    \begin{enumerate}
      \item Feedback can change the distribution of the total SFR-density over haloes of different masses (as seen in Figure~\ref{fig:rhosfrperdex}). Although the bulk of this effect likely results in a modification to the overall amplitude of the matter-SFR cross-correlation (e.g. as seen in the case of the \flammstar models), it also affects the small-scale behaviour, through the mass dependence of halo sizes.
      \item Feedback can change the distribution of star-forming regions within haloes (e.g. the distribution of satellite galaxies), effectively modifying the $\rho_{\rm SFR}(r|M)$ halo profile. This is a 1-halo effect that cannot be captured at large-scales by $\brsfr$.
      \item Feedback can also change the matter overdensity itself by redistributing gas within haloes \citep[e.g. ejecting part of it beyond the virial radius, ][]{1402.4461}.
    \end{enumerate}
    Therefore, to understand how different feedback models affect the statistics of the SFR-density field, we must study not only $P_{m\mathrm{SFR}}(k)$ (shown in the lower panels of Fig. \ref{fig:PSmSFR}), but also the matter power spectrum $P_{mm}(k)$ (upper panels).

    Focusing first on the different \flamfgas{\pm N} models, we observe small differences in the large-scale amplitude of $P_{m{\rm SFR}}(k)$, compatible with the changes in $\brsfr$ discussed in the previous section (with stronger feedback scenarios leading to marginally larger $\brsfr$ values). On small scales, correcting for the difference in the large-scale amplitude, we observe a suppression of power that becomes stronger with stronger feedback. The level of suppression is comparable to that experienced by the matter power spectrum itself, however. Thus, most of this effect may be explained by the impact of feedback on the matter overdensity itself, rather than causing a significant redistribution of the star-forming systems. Note, however, that AGN feedback is known to affect the distribution of satellite galaxies \citep{1310.7571,2412.08623,2512.04080}, and this is likely responsible for part of this suppression. 

    The \flammstar models show a different behaviour. The matter-SFR cross-spectrum is significantly more suppressed on large scales, in agreement with the lower $\brsfr$ recovered by these models, but this level of suppression is approximately the same on small scales, particularly at high redshifts. This is in agreement with the interpretation discussed in the previous sections: the main effect of the \flammstar models is an overall reduction in the SFR across all halo masses, and not to a modification in the spatial distribution of star-forming systems or in the SFR of specific halo mass ranges.
  \begin{figure*}
    \centering
    \includegraphics[width=0.7\textwidth]{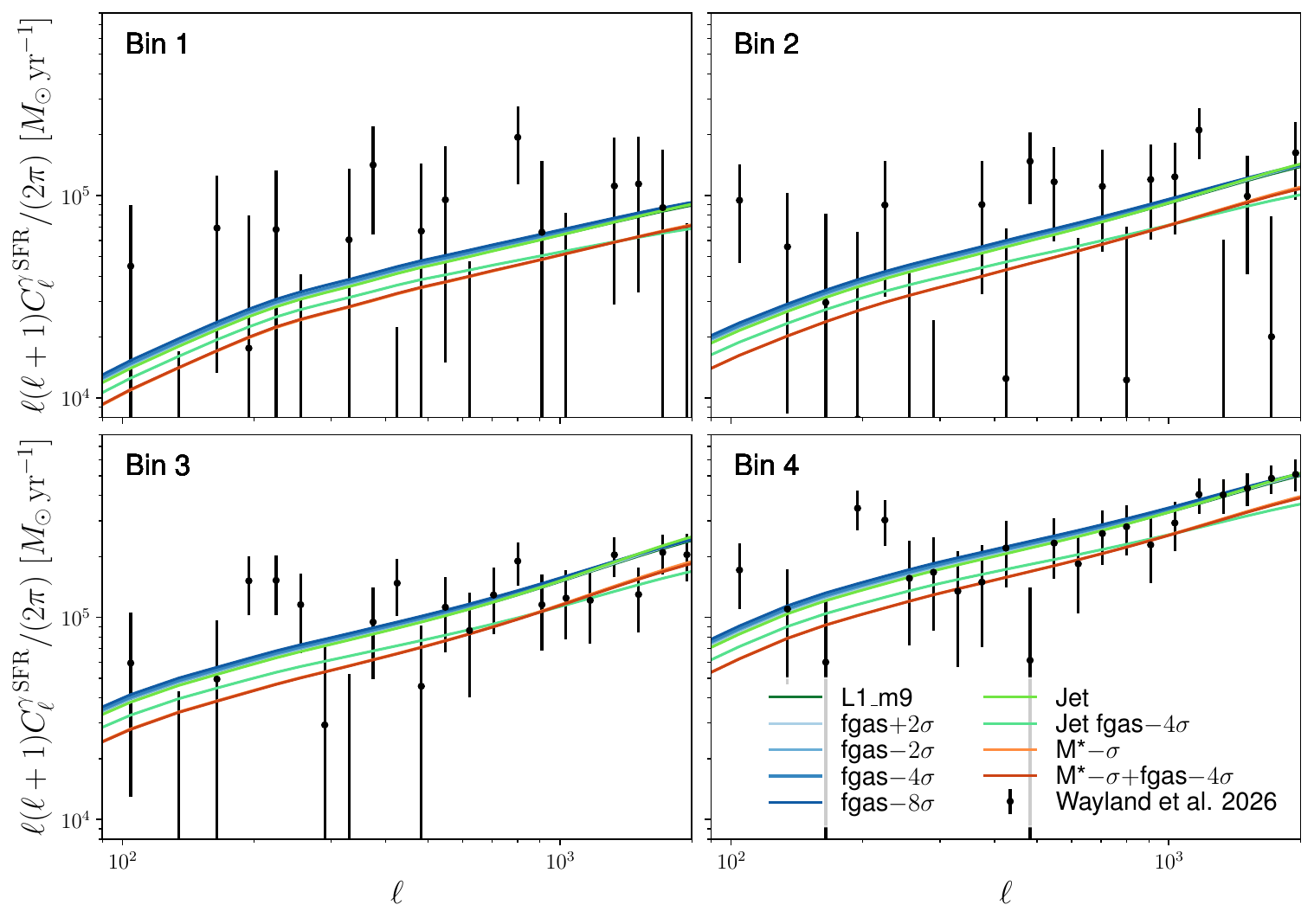}
    \caption{Angular power spectrum between cosmic shear and the CIB as a tracer of star formation. The black points with error bars correspond to measurements using shear data from DES Year-3 and multi-frequency \planck maps (details of this measurement can be found in \protect\cite{2609.31774}). The solid lines show the prediction for this quantity derived from the matter-$\rho_{\rm SFR}$ power spectrum measured in the different \flamingo simulations. Predictions are shown for models with different feedback models. Different thermal AGN models lead to almost identical predictions, whereas the strong jet-driven feedback, as well as strong stellar feedback models lead to a suppressed power spectrum. The data favours the fiducial \flamingo simulation over the latter two models at the $\sim 3\sigma$ level.}
    \label{fig:cl_wl_cib}
  \end{figure*}

    The two {\tt Jet} models show a more complex response, consistent with the redistribution of star formation across halo masses identified in Section~\ref{ssec:res.SFR_mass_rel}. Both kinetic-feedback models enhance the contribution of lower-mass haloes to the SFR-density, with a diverging behaviour at the high-mass tail (while \flamjet enhances it, \flamjethi suppresses it -- see Fig. \ref{fig:rhosfrperdex}). At $z=0.5$ the \flamjet model enhances both the matter power spectrum and the matter-SFR cross-power spectrum on non-linear scales, although the enhancement is stronger in the latter case. This likely due to a combination of two effects: the aforementioned enhancement of SFR in massive ($M_{200c}\gtrsim10^{13}\,M_\odot$) haloes, as well as a modification in the small-scale distribution of star-forming systems within those haloes. This agrees with the prediction from the \flamjethi simulation, which produces a large-scale suppression (consistent with the lower $\brsfr$ at low redshifts) and an even stronger suppression on small scales. This suppression is thus caused by the abundance and distribution of SFR systems in $\gtrsim10^{13}\,M_\odot$ haloes, where both simulations predict different contributions.  At $z=2.0$ this behaviour changes: $P_{m\mathrm{SFR}}(k)$ displays a suppression on small scales that is not reproduced in $P_{mm}(k)$ for both \flamjet and \flamjethi, stronger in the latter case. As we saw in Fig. \ref{fig:rhosfrperdex}, the main effect of the \flamjet model at high redshifts is the enhancement of SFR in low-mass haloes, with almost no change at higher masses with respect to the fiducial model. The \flamjethi follows the same behaviour at low masses while suppressing the high-mass contribution to $\rsfr$. The small-scale suppression observed is thus likely caused by the relative lack of clustering from star-forming systems in high-mass haloes compared to low-mass objects (i.e. it tracks the behaviour of the mean SFR-weighted halo mass shown in Fig. \ref{fig:Mh_SFR}). Modifications in the internal spatial distribution of star formation may also play a role in explaining this behaviour. Overall, the contrasting behaviour of the two {\tt Jet} models, compared with the relatively mild changes observed in the different \flamfgas{\pm N} simulations, suggests that the small-scale matter-SFR correlation may be sensitive to the details of the underlying AGN feedback model, at least within the calibration constraints used by \flamingo (namely the $z=0$ SMF). In the cases explored here, these details may be either the kinetic nature of feedback in the two {\tt Jet} models, or the different black-hole repositioning prescription used in these simulations.

  \subsection{CIB-shear cross-correlations}\label{ssec:res.cell_gamma_CIB}
  \begin{table*}
    \centering
    \caption{Rescaling amplitude $A_S$ of the shear-CIB power spectrum for the different \flamingo simulations and DES redshift bins. The fiducial \flamingo model, as well as all \flamfgas{\pm N} variants are in excellent agreement with the data, whereas the strong jets and the \flammstar models predict a cross-correlation that is $\sim25\%$ lower in amplitude, and are disfavoured at the $3.5\sigma$ level.}
    \label{tab:tab_Ashear}
    \begin{tabular}{|l|c|c|c|c|c|}
      \hline
      Simulation $(S)$ & Bin 1, $z_{\rm peak}=0.15$ & Bin 2, $z_{\rm peak}=0.25$ & Bin 3, $z_{\rm peak}=0.37$ & Bin 4, $z_{\rm peak}=0.47$ & Global fit\\
      \hline
      \flamfid & $0.53\pm0.32$ & $0.89\pm0.16$ & $0.93\pm0.08$ & $0.99\pm0.06$ & $0.95\pm0.05$\\
      \flamfgas{+2} & $0.53\pm0.32$ & $0.88\pm0.16$ & $0.93\pm0.08$ & $0.98\pm0.06$ & $0.94\pm0.05$\\
      \flamfgas{-2} & $0.52\pm0.32$ & $0.87\pm0.16$ & $0.91\pm0.08$ & $0.96\pm0.06$ & $0.93\pm0.05$\\
      \flamfgas{-4} & $0.52\pm0.31$ & $0.86\pm0.16$ & $0.91\pm0.08$ & $0.96\pm0.06$ & $0.92\pm0.05$\\
      \flamfgas{-8} & $0.50\pm0.31$ & $0.85\pm0.16$ & $0.90\pm0.08$ & $0.94\pm0.06$ & $0.91\pm0.05$\\
      \flamjet & $0.53\pm0.32$ & $0.88\pm0.16$ & $0.92\pm0.08$ & $0.98\pm0.06$ & $0.94\pm0.05$\\
      \flamjethi & $0.65\pm0.39$ & $1.14\pm0.21$ & $1.21\pm0.10$ & $1.27\pm0.08$ & $1.24\pm0.07$\\
      \flammstar & $0.67\pm0.41$ & $1.14\pm0.21$ & $1.21\pm0.10$ & $1.28\pm0.08$ & $1.23\pm0.07$\\
      \flammstarhi & $0.67\pm0.41$ & $1.14\pm0.21$ & $1.23\pm0.10$ & $1.29\pm0.08$ & $1.24\pm0.07$\\
      \flamls & $0.61\pm0.37$ & $1.00\pm0.19$ & $1.03\pm0.09$ & $1.10\pm0.07$ & $1.06\pm0.06$\\
      \flamlshi & $0.58\pm0.35$ & $0.96\pm0.18$ & $1.00\pm0.08$ & $1.05\pm0.06$ & $1.01\pm0.05$\\
      \hline
    \end{tabular}
  \end{table*}
  The differences observed in the matter-SFR power spectrum, presented in the previous section, manifest themselves directly in measurements of the cross-correlation between the CIB and cosmic shear data. Here we use the measurements of $C_\ell^{\gamma{\rm SFR}}\equiv \langle \gamma^E_{\ell m} \Sigma_{{\rm SFR},\ell m}^*\rangle$ presented in \cite{2609.31774}. These data improve on past measurements \citep[e.g. ][]{2209.05472} in different ways. First, they employed the latest \planck NPIPE frequency maps~\citep{Akrami2020planck}, as well as newer cosmic shear data from the third-year data release by the Dark Energy Survey \citep[DES-Y3, ][]{2011.03408,AbbottEtal2022}. Secondly, the analysis employed a novel component separation technique that reduces biases from correlated contaminants, such as the thermal Sunyaev-Zel'dovich effect. The cross-correlation signal is detected at high significance ($\sim20\sigma$).

  To estimate the predicted $C_\ell^{\gamma{\rm SFR}}$ from \flamingo, we begin by constructing an interpolating function for $P_{m{\rm SFR}}(k,z)$ from the power spectra measured in the simulations in the redshift interval $0\leq z\leq 3$. We then use the Core Cosmology Library \citep{1812.05995} to transform this interpolated 3D power spectrum into its 2D angular counterpart $C_\ell^{\gamma{\rm SFR}}$ using Eq. \ref{eq:cl_limber}, with the radial kernel for $\Sigma_{\rm SFR}$ given by $q_{\rm SFR}(\chi)\equiv4\pi\chi^2$ (according to Eq. \ref{eq:sigma_sfr}), and with the lensing kernel given by Eq. \ref{eq:q_gamma}. The resulting predictions are shown, together with the measurements of \citet{2609.31774} in Fig. \ref{fig:cl_wl_cib} for the four redshift bins of the DES-Y3 dataset. The cross-correlation is detected more significantly in the two highest redshift bins, where both the lensing and CIB signals are stronger. The prediction for the fiducial \flamingo simulation, as well as all the  \flamfgas{\pm N} variations, and the \flamjet model, all agree well with the measurements, while the simulations with stronger stellar feedback (\flammstar and \flammstarhi), as well as the simulation with strong jet-driven AGN feedback \flamjethi predict a $\sim 20$-$30\%$ lower cross-correlation on small scales. This reproduces the behaviour observed at the level of $P_{m{\rm SFR}}(k)$, with the latter three ``low-SFR'' models shown in Fig. \ref{fig:PSmSFR} consistently predicting a lower signal.

  To quantify the level of agreement with the data of a given simulation $S$, we rescale its theoretical prediction by a free multiplicative amplitude parameter $A_S$, and fit this simple model to the cross-correlation measurements to produce constraints on $A_S$. Disagreement with the data can then be quantified by the significance of the departure from the fiducial amplitude $A_S=1$. The results are shown in Table \ref{tab:tab_Ashear} for the different simulation and DES Y3 redshift bins. The result of a global fit to all redshift bins simultaneously is shown in the rightmost column. The measurement errors are too large in the first two bins to cleanly separate the different models, and all are compatible with $A_S=1$ at the 1-2$\sigma$ level. However, in the third and fourth redshift bins, we find that the three ``low-SFR'' simulations under-predict the measurement at the level of $\sim2.1\sigma$ and $\sim3.5\sigma$, respectively, while all other simulations are consistent with the data within $\sim1\sigma$. We thus find that current CIB-shear cross-correlations, in combinations with tomographic measurements of the bias-weighted SFR-density (see Fig. \ref{fig:rhoSFR}), disfavour models with strong stellar feedback, represented by the \flammstar and \flammstarhi simulations, as well as the strong jet-driven AGN feedback model in \flamjethi. The small-scale clustering information present in $C_\ell^{\gamma{\rm SFR}}$ is thus highly complementary to the tomographic $\brsfr$ measurements, which could not on their own distinguish between the fiducial feedback model and \flamjethi (see Fig. \ref{fig:rhoSFR}).

  \subsection{Dependence on cosmology}\label{ssec:res.cosmo}
    \begin{figure}
      \centering
      \includegraphics[width=\linewidth]{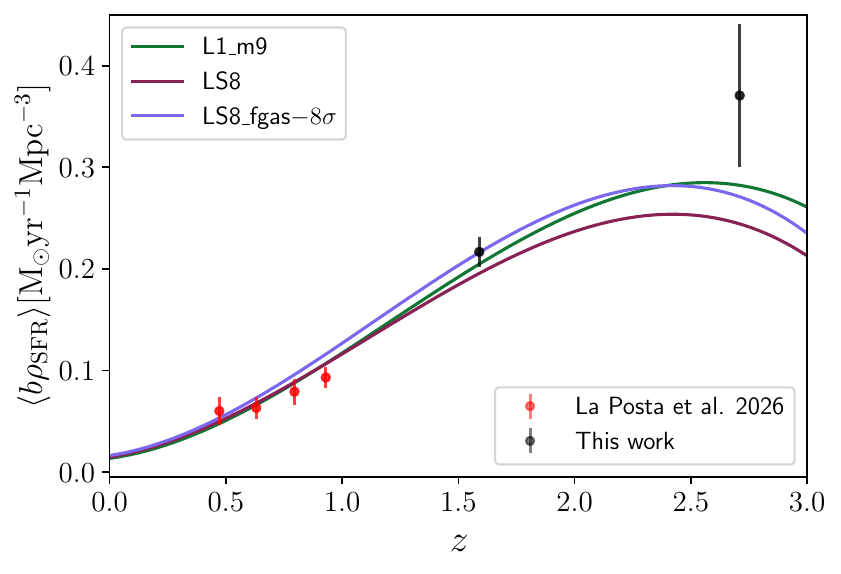}
      \includegraphics[width=\linewidth]{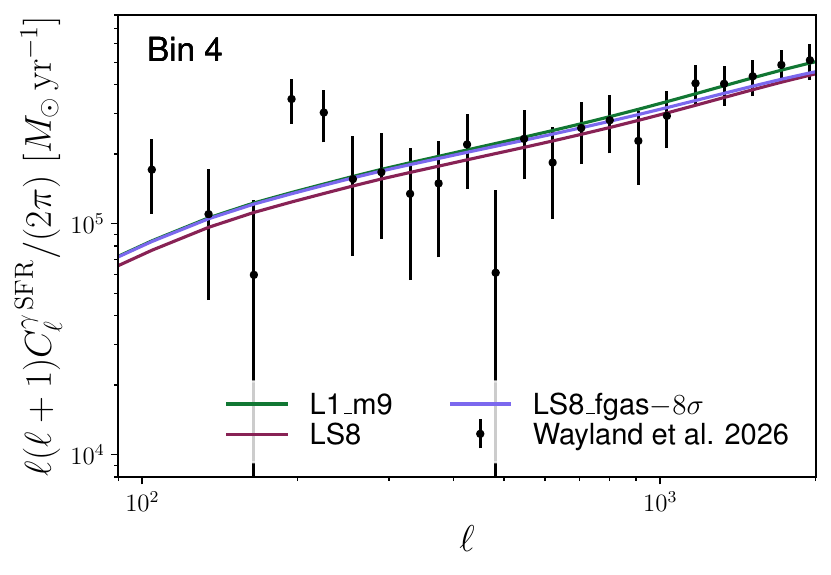}
      \caption{Measurements of $\brsfr$ and $C_\ell^{\gamma{\rm SFR}}$ (focusing on the fourth DES redshift bin), compared with the \flamingo predictions for the fiducial model (D3A cosmology, $S_8=0.815$), as well as simulations adopting the \flamls cosmological model, with a lower value of $S_8=0.766$. In the latter case, results are shown for the fiducial and strong AGN feedback models.}
      \label{fig:obs_cosmo}
    \end{figure}
    As we have seen in the previous sections, the \flamingo predictions for CIB cross-correlations in terms of star formation are largely constrained by the calibration to the observed $z=0$ SMF. Nevertheless, it is interesting to explore whether any residual dependence remains on the cosmological parameters describing the growth of the matter inhomogeneities which star-forming galaxies populate. Fig. \ref{fig:obs_cosmo} shows the measurements of $\brsfr$ and $C_\ell^{\gamma{\rm SFR}}$ (focusing on the fourth redshift bin, where the singal is highest) together with the \flamingo predictions for the fiducial \flamfid model as well as for simulations with a different set of cosmological parameters, close to the cosmological model preferred by \citet{2202.07440} (the \flamls model in the \flamingo nomenclature). Most importantly, this model features a lower value of the ``clumping'' parameter $S_8\equiv\sigma_8\sqrt{\Omega_m/0.3} = 0.766$ (compared with the baseline D3A cosmology having $S_8=0.815$). It is worth noting that the subgrid parameters in the \flamls simulations were not recalibrated with respect to those used in the corresponding simulations using the D3A cosmology, as both cosmologies were found to be sufficiently close to one another \citep{2018MNRAS.476.2999M, SchayeEtal2023}. Any differences we find here can therefore be fully ascribed to the change in cosmological parameters.
    
    We show predictions for two \flamls simulations, one with the fiducial strength for AGN and stellar feedback (simply labelled \flamls), and one with the strongest AGN feedback model (\flamlshi). For both observables we find that the differences between these models and the fiducial {\flamfid} are significantly smaller than those of the \flammstar and \flamjethi simulations discussed in previous section. As before, this result is driven by the calibration to the $z=0$ SMF. More in detail, the \flamls model predicts a lower value for $\brsfr$ above $z\sim1.5$. In turn, the \flamlshi model predicts a larger $\brsfr$, particularly in the high-redshift regime. This is in agreement with the dependence on the strength of thermal AGN feedback studied in Section \ref{ssec:res.tomography} (left panel in Fig. \ref{fig:rhoSFR}). Thus, CIB tomography at high redshifts is simultaneously sensitive to cosmology and AGN feedback, although current measurements are not sufficiently precise to place meaningful constraints on either effect.
    
    Turning to the cosmic shear cross-correlation, the \flamls model results in a very mild, and almost scale-independent suppression of the power spectrum. The \flamlshi model in turn raises the large-scale amplitude of the spectrum, matching the fiducial \flamingo prediction, while reproducing the same small-scale suppression of the \flamls simulation. Both effects are, in any case, very small, and indistinguishable from the fiducial prediction given current uncertainties. In terms of the $A_S$ parameter, the constraints obtained from all cross-correlation measurements are
    \begin{equation}
      A_{\tt LS8}=1.06\pm0.06,\hspace{12pt}
      A_{{\tt LS8\,fgas}-8\sigma}=1.01\pm0.05.
    \end{equation}
    Both are compatible with $1$ within $1\sigma$, and deviate from the fiducial prediction ($A_{\tt L1\_m9}=0.95\pm0.05$) by less than $2\sigma$. SFR-lensing cross-correlations are thus very weakly sensitive to the background cosmological parameters.

\section{Conclusions}\label{sec:conc}
  In this work, we have used the \flamingo cosmological hydrodynamical simulations to investigate how different implementations and calibrations of baryonic feedback affect the distribution and clustering of star formation, comparing their predictions for the statistics of the SFR-density field against CIB cross-correlation measurements.

  We find that the mass of haloes dominating the SFR-density remains remarkably stable across most feedback variations. The \flammstar models, which are calibrated to the observed $z=0$ SMF shifted to 0.14 dex lower stellar masses, primarily suppress the amount of star formation occurring around these halo masses, while variations in the cluster gas-fraction calibration produce comparatively small changes in the SFR distribution. The two {\tt Jet} kinetic AGN feedback models provide a more stark deviation, redistributing star formation across halo mass, increasing the relative contribution from lower-mass haloes, and suppressing the contribution from high masses when the feedback strength is raised. These results demonstrate that different feedback prescriptions can modify either the amount of star formation or the halo populations in which it occurs, although the result depends strongly on the observation these prescriptions have been calibrated on (critically, in the case of \flamingo, the $z=0$ SMF). We also report no significant cosmology dependence, which we believe once more to be due to the SMF calibration.

  These differences are reflected in the large-scale halo-bias-weighted SFR-density $\brsfr$. The \flamfgas{\pm N} variations remain close to the fiducial prediction, as do the ones with a different cosmology, while the \flammstar models show a systematically lower amplitude, consistent with their suppressed SFR-density per halo-mass bin. The two {\tt Jet} models instead predict an enhanced $\brsfr$ at high redshift, despite shifting star formation towards lower-mass haloes. This is due to the significant enhancement in the star formation occurring in these haloes, sufficient to overcome the reduction in halo bias. Current tomographic CIB measurements are broadly consistent with most simulations, although they strongly disfavour the high stellar feedback \flammstar models, particularly at intermediate redshifts ($z\sim1.5$). This implies a non-trivial consistency between observationally-inferred star-formation rates and stellar masses, and illustrates how CIB tomography can provide an independent test of feedback models away from the low-redshift observables on which they were calibrated. 

  On non-linear scales, the matter-SFR cross-power spectrum reveals additional sensitivity to baryonic physics. In this regime, the response depends not only on the amount and halo-mass distribution of star formation, but also on the redistribution of matter and the spatial relation between matter and star formation on halo scales. These differences propagate directly into the CIB-shear cross-correlation. The fiducial \flamfid, \flamfgas{\pm N} variations, \flamjet, and both \flamls models are consistent with the DES-Y3 measurements, whereas the \flammstar, \flammstarhi, and \flamjethi models are disfavoured at the $\sim3.5\sigma$ level, driven by the small-scale cross-correlation signal at the highest redshift bin. Current CIB-shear measurements therefore already provide discriminatory power between feedback prescriptions, disfavouring the models calibrated to the $z=0$ SMF shifted to lower masses, and strong jet-driven AGN feedback model. In the latter case, the increased quenching that drives the disagreement with the data may be due to the distinct black-hole repositioning approach implemented in this model.

  Overall, our results show that CIB cross-correlations probe more than the integrated cosmic star-formation history: they contain information about which haloes host star formation and how star-forming galaxies are spatially connected to the matter distribution. Combined with hydrodynamical simulations, these measurements therefore provide a promising route for testing the baryonic processes that regulate galaxy formation across cosmic time. On the simulation side, the COLIBRE suite of simulations \citep{2026MNRAS.548ag375S, 2026MNRAS.548ag300C} provide higher resolution and small-scale fidelity than \flamingo, and are able to reproduce the observed evolution of the SMF and the SFR-density \citep{2509.07960}. On the observational side, new weak lensing data from ongoing surveys, such as Euclid and the Rubin Observatory, in combination with more sensitive and higher-resolution sub-millimetre and far-infrared data from experiments such as the Simons Observatory and the Fred Young Submillimeter Telescope, will soon allow us to characterise the relation between matter and star formation on small scales with much higher precision. Likewise, deeper galaxy samples, (e.g. Lyman-break galaxies from the Rubin Observatory) will allow us to reconstruct the star formation history through tomography to higher redshifts ($3\lesssim z\lesssim 5$). With these more sensitive datasets,  most of the approximations adopted in this analysis will need to be reassessed. This includes the infrared spectral templates used to reconstruct the CIB component from multi-frequency cross-correlations \citep{2026arXiv260304269L,2609.31774}, and non-linear contributions to the relationship between infrared luminosity and SFR.  Importantly, our analysis has ignored the impact of dust heating by AGN emission, which could provide a novel avenue to constrain the physics of AGN feedback. Ultimately, reproducing the raw measurements (e.g. multi-frequency cross-correlations) using simulation-based products \citep[see ][for an example using \flamingo]{2512.09891} will enable a simultaneous test of these approximations. We leave such studies for future work.


\section*{Acknowledgements}
  We would like to thank Julien Devriendt, Anna Ivleva, Abhishek Maniyar, Adrianne Slyz, and Imogen Whittam for useful discussions.

  SMM, and DA acknowledge support from the Beecroft Trust. ALP and DA are supported by the Science and Technology Facilities Council (STFC) under grants with reference UKRI1164 and ST/W000903/1. RD acknowledges support from the Oxford Astrophysics Summer Internship programme. AW is supported by a Science and Technology Facilities Council (STFC) studentship. 
  
  This work used the DiRAC@Durham facility managed by the Institute for Computational Cosmology on behalf of the STFC DiRAC HPC Facility (\url{www.dirac.ac.uk}). The equipment was funded by BEIS capital funding via STFC capital grants ST/K00042X/1, ST/P002293/1, ST/R002371/1 and ST/S002502/1, Durham University and STFC operations grant ST/R000832/1. DiRAC is part of the National e-Infrastructure.

  This work was also supported by the Science and Technology Facilities Council (grant number ST/Y002733/1).

\section*{Data Availability}
  Any simulation-based predictions used for this paper and not already available in the \flamingo data release\footnote{\url{https://dataweb.cosma.dur.ac.uk:8443/flamingo/introduction.html}}, as well as the data used to infer the high-redshift measurements of $\brsfr$ shown in Fig. \ref{fig:rhoSFR}, can be provided upon request.



\bibliographystyle{mnras}
\bibliography{PAPER_v1} 



\appendix

\bsp	
\label{lastpage}
\end{document}